\documentclass[aps,prb,twocolumn]{revtex4}   
\usepackage{graphicx}

\usepackage{amsmath}   
\usepackage{subfigure} 
\usepackage{amssymb}    
\usepackage{textcomp}
\usepackage{graphicx}   
\usepackage{dsfont}
\usepackage{array}
\usepackage{epstopdf}

\begin{document}
\title{Accelerated Molecular Dynamics through stochastic iterations to strengthen yield of path hopping over upper states (SISYPHUS)}
\author{Pratyush Tiwary}
\author{Axel van de Walle}
\affiliation{School of Engineering, Brown University, Providence, Rhode Island 02912, USA}
\date{\today}

\begin{abstract}
We present a new method, called SISYPHUS (Stochastic Iterations to Strengthen Yield of Path Hopping over Upper States), for extending accessible time-scales in atomistic simulations.
The method proceeds by separating phase space into basins, and transition regions between the basins based on a general collective variable (CV) criterion.
The transition regions are treated via traditional molecular dynamics (MD) while Monte Carlo (MC) methods are used to (i) estimate the expected time spent in each basin and (ii) thermalize the system between two MD episodes. In particular, an efficient adiabatic switching based scheme is used to estimate the time spent inside the basins. The method offers various advantages over existing approaches in terms of (i) providing an accurate real time scale, (ii) avoiding reliance on harmonic transition state theory and (iii) avoiding the need to enumerate all possible transition events.
Applications of SISYPHUS to low temperature vacancy diffusion in BCC Ta and adatom island ripening in FCC Al are presented. A new CV appropriate for such condensed phases, especially for transitions involving collective motions of several atoms, is also introduced.
\end{abstract}
\maketitle

Achieving usefully long timescales (seconds or longer) in atomistic simulations of materials is a problem of great interest, and the search for a practical and general solution has generated intense activity in the field over last several decades\cite{metadynamics,hyperdynamics,mcmd,abc,ontheflykmc,art,deshaw,kappa,blas_science} (see \cite{voter_review,laio_review,perez_review} 
for excellent reviews of these and other efforts). The problem arises because, as the system moves from one energy basin to another through infrequent rare events, it stays trapped in some energy basin for extended periods of time. Along with the small time steps (on the order of femtoseconds) needed for the total energy staying conserved, this severely restricts the time scales accessible in MD simulations and also leads to limited phase-space exploration.

Although many methods exist to increase the rate of rare events and efficiently explore the rugged free energy landscape, a frequent limitation is the inability to efficiently obtain an accurate estimate of the ``real'' time scale of the simulation in general physical systems. Existing schemes to achieve this either
\begin{enumerate}
\item require cataloguing all possible transitions paths out of given basin\cite{ontheflykmc,art,abc}, which can be computationally prohibitive in low-symmetry systems, or
\item rely on harmonic approximations to the system's energy surface\cite{tad}, or 
\item involve computing averages that converge slowly, especially for large system sizes, because they involve exponentials of the system's total energy.\cite{hyperdynamics, jarzynski_rare}
\end{enumerate}

In this letter, we propose a new mixed Monte Carlo-Molecular Dynamics method that uses a collective variable (CV) $\chi$ solely to discriminate between basins and transition regions, thus placing very weak requirements on the choice of CV, less stringent than what required in metadynamics methods\footnote{In Metadynamics, the CV should distinguish between initial basin, final basin and transition regions. Our CV however is allowed (but not required) to take the same value in two basins.}. The method still provides the system's dynamics via conventional atomic coordinates and thus provides more detailed information than the coarse grained dynamics typically provided by Metadynamics methods\cite{metadynamics,metadynamics_energy}. The method also provides an accurate real time scale that does not deteriorate with system size and that does not rely on harmonicity assumptions. There is no need to construct \textit{a priori} or \textit{in situ} a catalog of possible transition mechanisms. The method is specially suited for exploiting massive parallel computing.

The proposed algorithm generalizes our previous work\cite{mcmd} along multiple dimensions. First, we use a general CV $\chi$ (instead of the system's potential energy) to discriminate between basins and transition regions and propose a novel type of CV suitable for this purpose in condensed phases. (Recently proposed dimensionality reduction algorithms\cite{sketchmap1,sketchmap2} that discover CV automatically could be used as well.) Second, we introduce an adiabatic switching scheme to efficiently calculate the real time spent inside wells. Finally, we use a more robust criterion to determine when the system has been trapped in a basin for sufficiently long time to have equilibrated therein.

Let the system be characterized by position $x=(r_1,\ldots ,r_N)$ and velocity $v=(v_1,\ldots ,v_N)$. The CV $\chi$ is a function of $x$ (we give an example of such a function later in this letter). For a user-specified cut-off value $\chi_{cut}$, we define the basins, or wells, \textbf{W} in this $\chi$-space as a set of connected states for which $\chi < \chi_{cut}$. In the wells, the method does not follow the system's exact trajectory in phase space, but instead provides the expected amount of time spent in each well. In contrast, the $\chi\geq\chi_{cut}$ region of the phase space contains the interesting but infrequently occurring events whose dynamics is fully described. The choice of $\chi_{cut}$ thus specifies the level of details one wishes to retain in the simulation. Large values of $\chi_{cut}$ may cause wells to merge
and limit the ability to resolve the precise dynamics of some events. The method is still formally correct but the definition of the wells \textbf{W} may not have an an obvious physical meaning. Nevertheless, the method is very robust to changes in $\chi_{cut}$ that do not change the topology of the well structure (we will provide numerical evidence of this fact).

When the CV $\chi$ of the system is above the threshold $\chi_{cut}$, the system evolves via MD according to its true Hamiltonian with constant-pressure (or volume) or constant-temperature (or energy) ensemble as entailed by the simulation. When $\chi < \chi_{cut}$, the algorithm continues performing MD until either (i) $\chi \geq \chi_{cut}$ (in which case the system is considered to have exited the well and standard MD continues as described above) or (ii) a time equal to the system's decorrelation time $\tau_{c}$ has elapsed. In that latter case, the system is considered to have been trapped in the well for long enough that it has reached a local thermodynamic equilibrium. This criterion is similar to the one used in other methods \cite{parrep,kappa} to define a transition event. At this time, we launch a MC simulation (called MC$a$) whose aim is to generate a new random starting position at the well's boundary to initialize the next MD episode. MC$a$ is run for long enough that the system loses memory of how it entered the well and visits the boundary of the well a few times. MD is restarted with the position $x$ where the system last visited the well's boundary and with velocities $v$ drawn from a truncated Maxwell-Boltzmann distribution conditional on $v\cdot\nabla \chi(x)>0$ (i.e. we only consider velocities in the half-space pointing outwards of the well).

In parallel to the first MC (MC$a$) we perform a second MC run (called MC$b$) that calculates the expected time $t_{\textbf{W}}$ the system would have spent inside the well.
This separation of two MC runs makes our algorithm extremely parallelizable and especially amenable to be used on loosely coupled cluster of computers. We can launch as many MC$b$ runs as we have processors available. These runs do not need to communicate with each other, and because of the system's ergodicity, we can make a quick estimate of the quantity $t_{\textbf{W}}$ by averaging over these independent runs. This parallelization is even simpler than for the Parallel Replica method\cite{parrep}.

Before we describe how we calculate the time the system would have spent in whichever well it visited, we describe specifically how MC$a$ is implemented. We define the boundary of the well \textbf{W} with thickness $w$:
\begin{equation}
\label{eq:lid}
S_{\textbf{W}} = \{x =(r_1,...,r_N) : |\chi(x) - \chi_{cut}| < w  \}
\end{equation}
Trying to visit $S_{\textbf{W}}$ with an unbiased potential would be of no avail, since we will rarely visit states in $S_\textbf{W}$. We thus do MC with a biased potential $V^{*}(x)$ defined as follows:
\begin{eqnarray}
\label{eq:V_bias}
	V^{*}(x)  &=& V(x) 
	+  \left\{ 
\begin{array}{l l}
\infty & \quad {\chi(x) \geq \chi_{cut} } \\
V_0\left ( {{\chi_{cut} - \chi}\over {\chi_{cut}}}\right)^m & \quad {\chi(x) < \chi_{cut}}\\ \end{array} \right.
  \end{eqnarray}

Per Eq. (\ref{eq:V_bias}), MC$a$ never visits the outside of the well \textbf{W} and biases states with a penalty function that increases with their depth inside the well. Note that the bias is zero at the well boundary, which is important to obtain the correct sampling distribution of the boundary\cite{hyperdynamics}. The bias inside the well changes the fraction of time spent at the boundary but not the ratio of the times spent at any two points of the boundary.
The parameter $m$ determines how sharp the boundary of the well is (a value around 0.5 is found to perform well in practice). $V_0$ is kept around the standard deviation in potential energy of the system. As we demonstrate numerically later, the algorithm is very robust with respect to choice of parameters in Eq. (\ref{eq:V_bias}).

Having described a way to accelerate the exploration of various wells, we now turn to the question of calculating (via a Monte Carlo labelled MC$b$) the expected time the system would have spent inside well \textbf{W} if there was no acceleration of the dynamics. This time, denoted $t_{\textbf{W}}$, can be calculated as the reciprocal of the flux of states exiting the well:
\begin{equation}
\label{eq:time}
t_\textbf{W} = \lim_{w \rightarrow 0} ( \langle  {\overline{v}\over w} \; 1(x \in \textbf{S$_w$}) \rangle )^{-1}
\end{equation}
where the average $\langle\cdots\rangle$ is taken over $x$ drawn from the well \textbf{W} with a probability density proportional to $e^{-V(x)/(k_B T)}$. $k_B$ is Boltzmann's constant, T is the temperature and $1(A)$ equals 1 if the event $A$ is true and 0 otherwise. $\overline{v}$ denotes the mean projection of a Maxwell-Boltzmann-distributed velocity along the unit vector $u$ parallel to $\nabla \chi(x)$, conditional on $v.\nabla \chi(x)<0$. When all atoms have the same mass $m$, $\overline{v} =\sqrt{k_B T/{2 \pi m}}$ (a general expression can be found in Ref. \onlinecite{mcmd}). 

\begin{figure}
 \includegraphics[width=80mm]{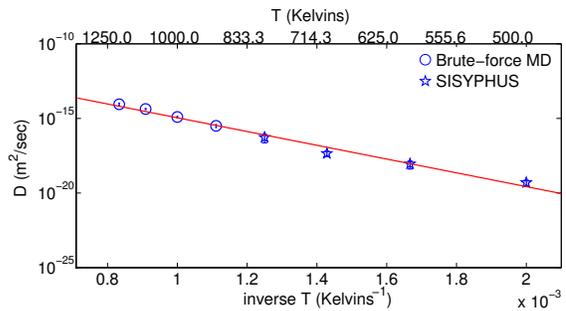}
 \caption[2]
{Diffusion constant for vacancy diffusion in Ta at various temperatures as through brute-force MD (circles) and SISYPHUS (stars). Errorbars (roughly same as marker size) are also provided as obtained over 16 independent runs.}
\label{fig:diffusion}
\end{figure}

\begin{figure*}[htp]
 \includegraphics[width=58mm]{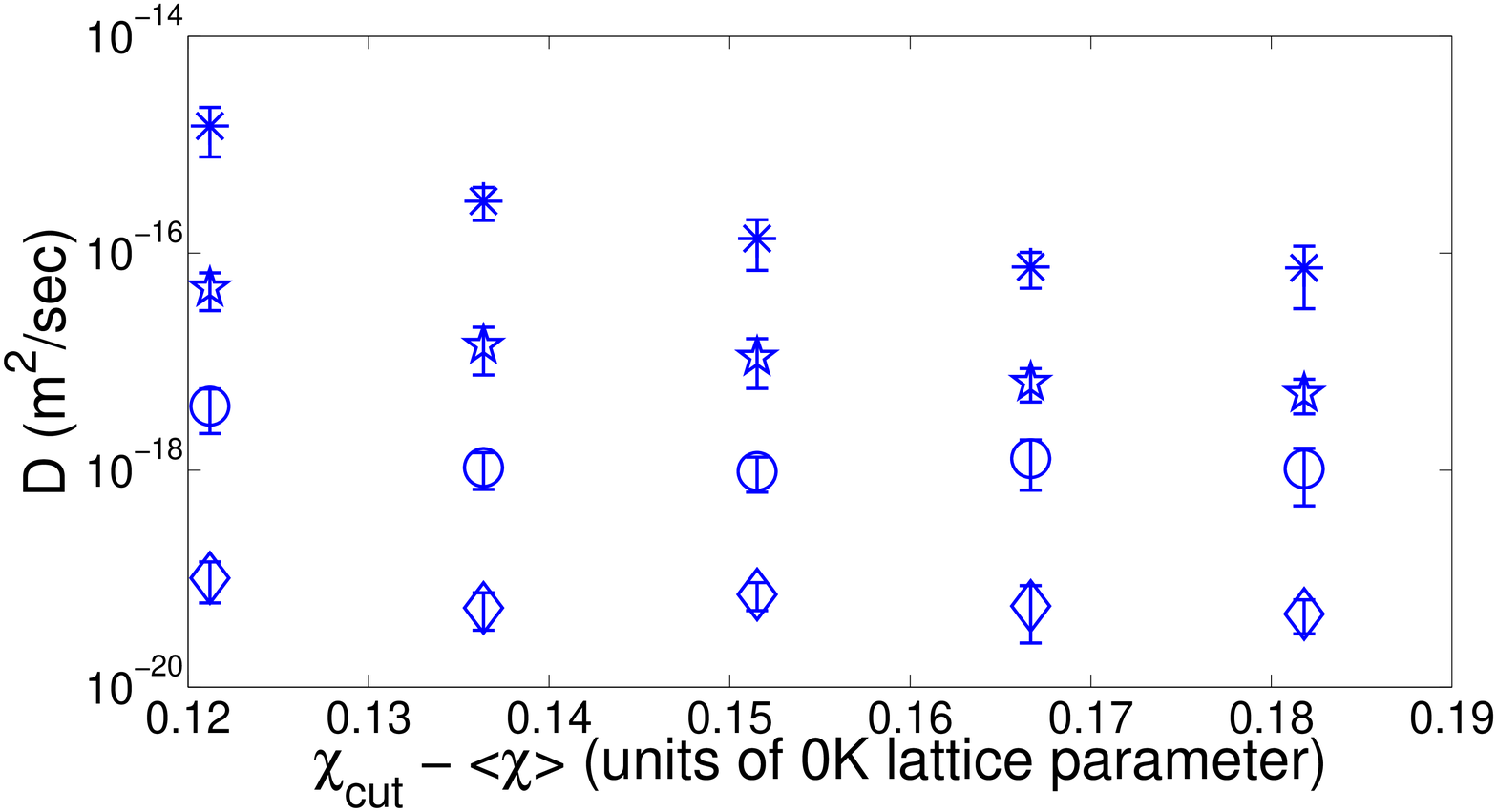}
 \includegraphics[width=58mm]{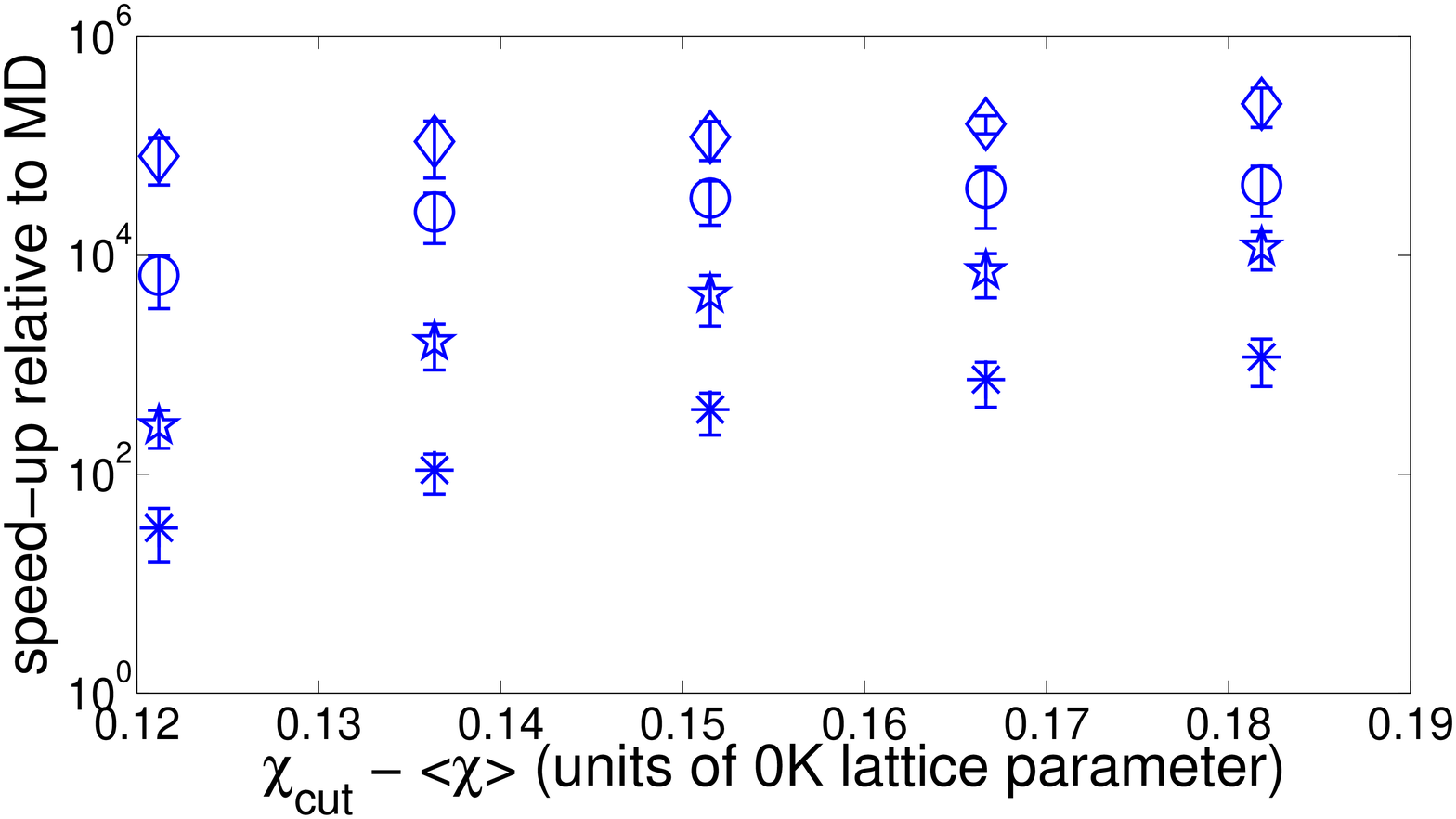}
 \includegraphics[width=58mm]{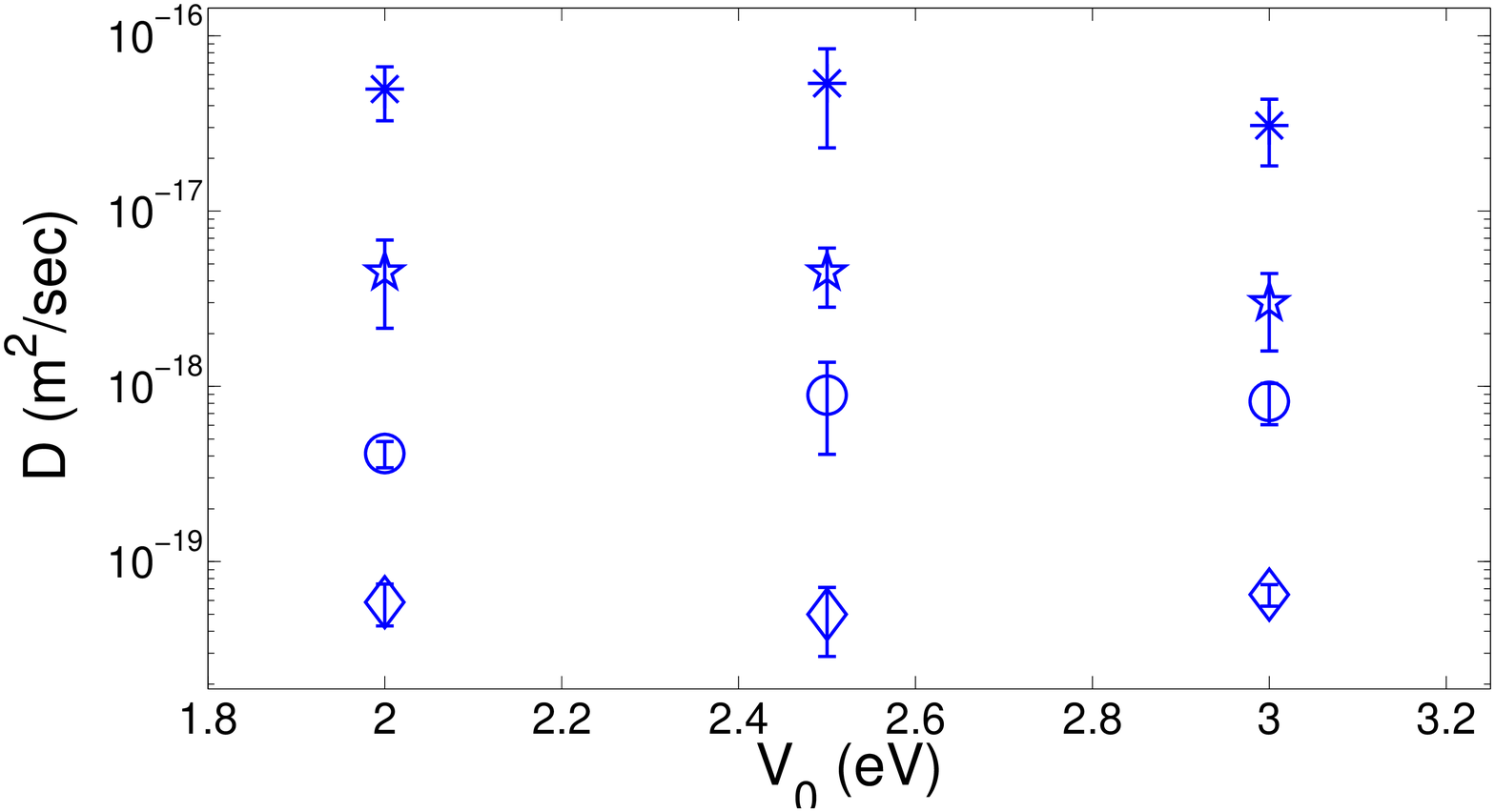}
 \caption[2]
{(Top) Insensitivity of dynamics to choice of $\chi_{cut}$ (relative to average $\chi$ at that temperature) for vacancy diffusion in Ta across temperatures. $\diamond$, \textopenbullet,$\star$ and $\ast$ denote 500,600,700 and 800K temperatures respectively. (Centre) Corresponding speed-ups relative to physical time achieved in brute-force MD run in the same wall-clock time. (Bottom) Insensitivity of dynamics to choice of $V_0$. Errorbars over 16 independent runs for each data point.}
\label{fig:chicut_robustness}
\end{figure*}

A straightforward implementation of Eq. (\ref{eq:time}) will again suffer from a rare event problem. Even an importance sampling scheme, as suggested in Ref. \onlinecite{mcmd} is not very efficient. Consider what happens if we used the biased potential as defined in Eq. (\ref{eq:V_bias}) to calculate $t_{\textbf{W}}$:
\begin{equation}
\label{eq:t_bias}
t_\textbf{W} = \lim_{w \rightarrow 0} \frac{  \langle  e^{-{{\beta}}(V(x)-V^{*}(x))} \rangle^* }
{\langle {\overline{v}\over w} {e^{-{{\beta}}(V(x)-V^{*}(x))} }1(x \in \textbf{S$_w$}) \rangle^* }
\end{equation}
where $\langle \cdots \rangle*$ denote expectations taken under a density proportional to $e^{-\beta V^*(x)}$, in which $\beta$ is $1/(k_B T)$. This approach is exact in the limit of an ensemble average, but there is a fundamental trade-off that limits its usefulness: A large bias $V^{*}(x)-V(x)$ leads to a more rapid convergence of the denominator (due to an increased sampling rate of the boundary) but a slower convergence of the numerator (due to an increase in $V^{*}(x)-V(x)$).

To avoid this problem, we now propose a technique that bears some resemblance to adiabatic switching methods\cite{frenkelsmit,jarzynski}, in which the system is continuously, \emph{adiabatically} switched from $V\left( x\right) $ (the true potential) to $V^{*}(x)$ (identical to the potential used in MC$a$). Let $\hat{V}\left( x,\alpha \right) $ smoothly interpolate between $\hat{V}%
\left( x,0\right) \equiv V\left( x\right) $ and $\hat{V}%
\left( x,1\right) \equiv V^{*}(x)$. Then we can express the ensemble average in Eq.(\ref{eq:time}) as below (working in terms of rate $t_\textbf{W}^{-1}$):%
\begin{eqnarray}
\label{eq:adb1}
t_\textbf{W}^{-1} &=&\lim_{w \rightarrow 0}\frac{\int {\bar{v}\over{w}} 1\left( x\in S_{w}\right)
e^{-\beta \hat{V}\left( x,0\right) }dx}{\int e^{-\beta \hat{V}\left(
x,0\right) }dx} \nonumber \\
&\equiv &\lim_{w \rightarrow 0} \left\langle {{\overline{v} 1\left( x\in S_{w}\right)}\over{w}} e^{-\beta
\left( \hat{V}\left( x,0\right) -\hat{V}\left( x,1\right) \right)
}\right\rangle _{1}R
\end{eqnarray}%
where $dx$ denotes a differential volume in 3-N dimensional configuration space for N particles, the integration being performed over entire configuration space within the well \textbf{W} and the expected value $\left\langle \cdots \right\rangle _{\alpha }$ in Eq.(\ref{eq:adb1})  is defined by
\begin{equation}
\label{eq:adb2}
\left\langle \cdots \right\rangle _{\alpha }=\frac{\int\left( \cdots
\right) e^{-\beta \hat{V}\left( x,\alpha \right) }dx}{\int e^{-\beta \hat{%
V}\left( x,\alpha \right) }dx}
\end{equation}
Below we define the term $R$ in Eq.(\ref{eq:adb1}) and re-express it in a computationally tractable form (see Supplemental Materials (SM) for a more detailed derivation):
\begin{eqnarray}
\label{eq:adb3}
R &=&\frac{\int e^{-\beta \hat{V}\left( x,1\right) }dx}{\int e^{-\beta 
\hat{V}\left( x,0\right) }dx} = \exp \left( -\beta \int_{0}^{1}\left\langle \frac{\partial \hat{V}\left(
x,\alpha \right) }{\partial \alpha }\right\rangle _{\alpha }d\alpha \right) \nonumber \\
\end{eqnarray}

We pick a linear switching scheme for $\hat{V}(x,\alpha)$, i.e. an interpolation scheme between $\hat{V}(x,0)$ and $\hat{V}(x,1)$:
\begin{equation}
\label{eq:switch}
\hat{V}(x,\alpha) = (1-\alpha)V(x) + \alpha V^{*}(x) 
\end{equation}
We now make a few observations regarding Eq.(\ref{eq:adb1}). It involves 2 parts. The first is $\lim_{w \rightarrow 0}\left\langle{{\overline{v} 1\left( x\in S_{w}\right)}\over{w}} e^{-\beta\left( \hat{V}\left( x,0\right) -\hat{V}\left( x,1\right) \right)}\right\rangle _{1}$. This is nonzero only when $x\in S_{w}$, and whenever it is nonzero, the difference $\hat{V}\left( x,0\right) -\hat{V}\left( x,1\right)$ is very small(see Eqs.(\ref{eq:lid}-\ref{eq:V_bias})). Since this average is calculated with the maximally biased potential $\hat{V}\left( x,1\right)$, the boundary $x\in S_{w}$ is visited frequently, and thus the first term in Eq.(\ref{eq:adb1}) can be evaluated very quickly. The second part in Eq.(\ref{eq:adb1}) is $R$, where the average $\langle \frac{\partial \hat{V}(x,\alpha) }{\partial \alpha }\rangle _{\alpha } = \langle{V^{*}(x) - V(x)}\rangle_{\alpha}$ does not contain any exponentials that could cause a slow convergence. In SM, we prove rigorously that for this switching scheme and for the choice of biasing potential in Eq.(\ref{eq:V_bias}) we can use a non-uniform grid to evaluate $R$ which can be made finer as $\alpha\to0$ but kept coarse for larger $\alpha$, leading to further computational efficiency.

\begin{figure*}
\subfigure[1 adatom (before)]{
\includegraphics[scale=0.1]{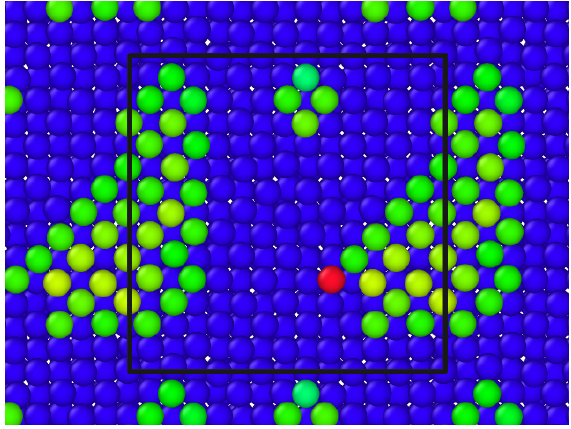}}
\subfigure[1 adatom (after)]{
\includegraphics[scale=0.1]{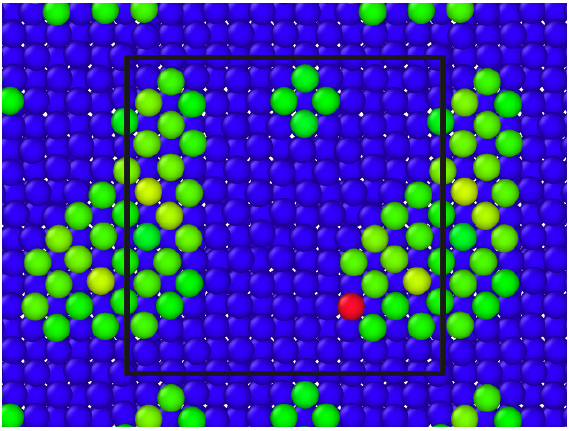}}
\subfigure[2 adatoms (before)]{
\includegraphics[scale=0.1]{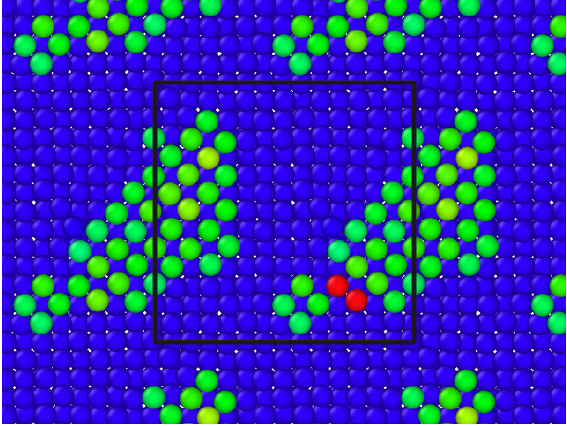}}
\subfigure[2 adatoms (after)]{
\includegraphics[scale=0.1]{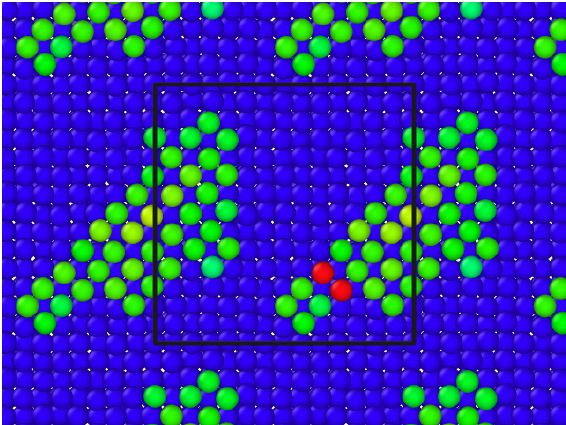}}
\subfigure[Ad \& substrate(before)]{
\includegraphics[scale=0.1]{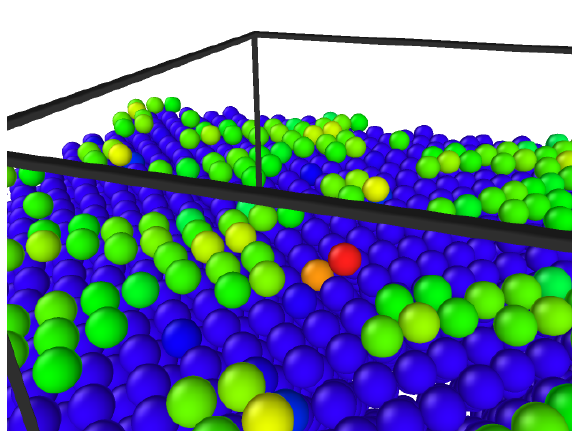}}
\subfigure[Ad \& substrate(after)]{
\includegraphics[scale=0.1]{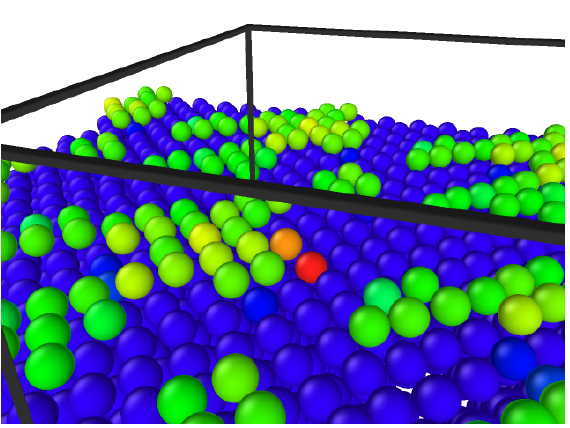}}
\subfigure[3 adatoms (before)]{
\includegraphics[scale=0.1]{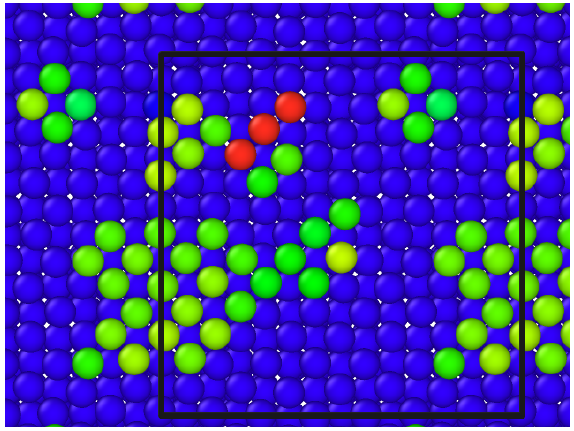}}
\subfigure[3 adatoms (after)]{
\includegraphics[scale=0.1]{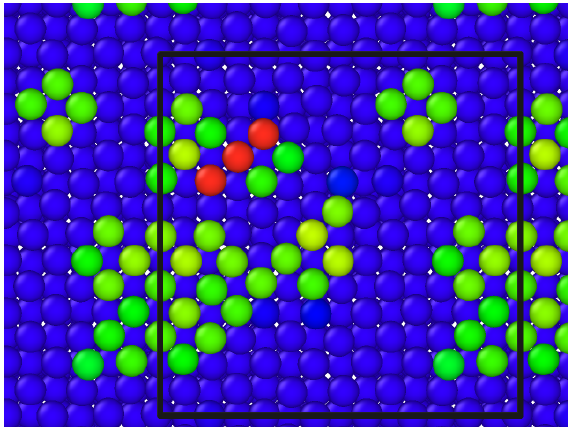}}
\caption{\label{fig:mechanisms} Mechanisms seen by SISYPHUS for adatom movement on Al(001) surface. Atoms colored per z-coordinate. Blue/orange = substrate atom, green/red=adatom, red/orange = atom with maximum movement. Solid lines are periodic boundaries.}
\end{figure*}

For solid-state systems where bond-breaking is the dominant mechanism of interest, we take $\chi$ to be the bond distortion function (BDF), defined below for a N-particle system:
\begin{equation}
\label{eq:bdf}
\chi(x) = a_0 \left (\sum_{\forall i,j \textrm{ } r_{ij}=|r_i-r_j|}  | {  {  r_{ij}-r_{ij}^{eq}  } \over{r_{ij}^{eq}} }  |^p \right ) ^{1/p}
\end{equation}
where $p>1$ and $a_0$ is the 0 Kelvin lattice parameter. For each bond, $r_{ij}^{eq}$ is the equilibrium bond length that can be obtained by a few conjugate gradient steps each time a transition is detected. In the limit that $p \to \infty$, the BDF, which is a p-norm over fractional bond distortions, approaches the maximum norm, i.e. the strain (times $a_0$) in the maximally strained bond. For $p \to \infty$ we thus recover the so-called bond-boost function\cite{bondboost}. We pick $p$ around 8-12 for the systems studied in this letter. Not taking $p  = \infty$ allows us to treat on a similar footing cases where (a) one bond is distorted by a large amount, or (b) several bonds are collectively distorted by a significant amount which is however less than the amount in (a). As soon as either (a) or (b) happens, the BDF detects it through a spike and thus we can then switch back to doing completely unbiased MD. This way we can treat transition mechanisms involving small but concerted and collective motion of several atoms (see Fig. \ref{fig:mechanisms} g-h)) - for example in glasses, shear transformation zones\cite{stz} involve several atoms moving displacing together by a small amount.

We now describe applications of the algorithm that validate it and demonstrate its insensitivity to choice of parameters. The first example is vacancy assisted lattice diffusion at low temperatures in BCC Tantalum. Lattice diffusion at low temperatures is a problem important in a spectrum of sciences from Materials Science to Geology\cite{defectsmaterials,epsl}, but is beyond the time scales one can access in current MD simulations, requiring times longer than milliseconds\cite{mendelev,mcmd}. We describe the parameters for the MD part\cite{ackland} of this simulation in SM. In Fig.\ref{fig:diffusion} we demonstrate how the diffusion constant through brute-force MD and SISYPHUS for vacancy assisted diffusion lie on the same Arrhenius plot, giving an Arrhenius-type activation energy of 0.9(+/-0.1) eV (in rough agreement with Harmonic Transition State Theory(HTST) calculation of 1.1 eV). Fig.\ref{fig:chicut_robustness}(top) demonstrates the lack of sensitivity of the dynamics to what $\chi_{cut}$ and $V_0$ values we pick. As expected the speed-up is higher for higher $\chi_{cut}$ (Fig.\ref{fig:chicut_robustness}(middle)). At higher temperature we find slightly higher sensitivity to $\chi_{cut}$ (still within order of magnitude accuracy). This is because for a low $\chi_{cut}$, the $t_{\textbf{W}}$ values becomes closer to time $\tau_{c}$ the system takes to equilibrate in a well. Insensitivity to $V_0$ values can be seen from Fig.\ref{fig:chicut_robustness}(bottom). A smaller $V_0$ leads to slower sampling in Eq.\ref{eq:t_bias}, and MC$a$ also takes longer to converge. With too high a $V_0$ however, the periphery of the well \textbf{W} can be too steep and one might again face sampling issues since the system can be trapped in some regions of the well boundary. In SM, we provide a back-of-the-envelope estimate of how to pick an optimum $V_0$ for a given system.

For our second application, we studied the room-temperature dynamics of Al adatoms on Al (001) thin film (625 atoms, roughly 3 times larger than the Ta vacancy diffusion example). We picked this problem because firstly, from a technological perspective, it is of immense importance for fabrication processes in nanoscale devices involving growth of thin films from deposited adatoms\cite{thinfilm_science,fichthorn_prl}. This problem is very interesting from a theoretical perspective too, given that it is an inherently non-equilibrium phenomenon dictated by the interplay between kinetics and thermodynamics. Being able to model and control the growth and properties of such films is hugely desirable -  the time-scales needed are however far beyond MD. Accurate 0 Kelvin saddle point search methods\cite{dimer,ontheflykmc} have shown the existence of a large number of transition mechanisms with low and similar activation energies (smaller than 0.4eV). Such low lying barriers can be hard to deal with in most accelerated MD methods\cite{voter_review}. On-the-fly KMC\cite{ontheflykmc} calculations have been used previously used to get rough estimate of the time-scale for adatom island ripening that we can compare SISYPHUS with. In SM we provide details of the simulation parameters for this system.

\begin{figure}
\subfigure[ t=0 ns (0 eV)]{
\includegraphics[scale=0.118]{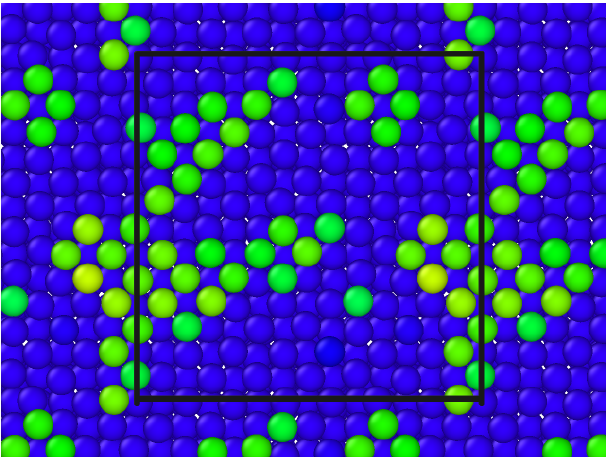}}
\subfigure[ t=1 $\mu$s (-1.6 eV)]{
\includegraphics[scale=0.118]{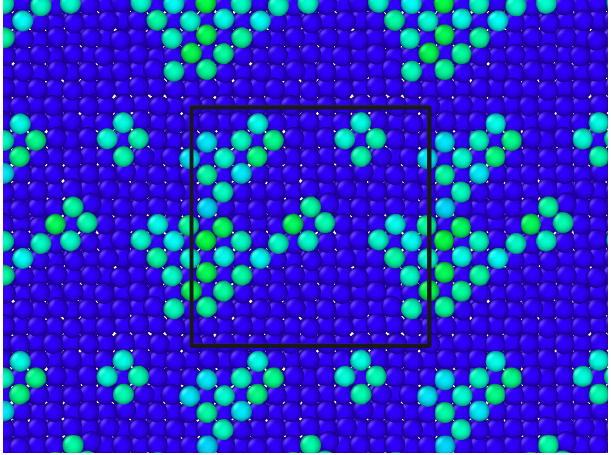}}
\subfigure[ t=2 ms (-1.8 eV)]{
\includegraphics[scale=0.112]{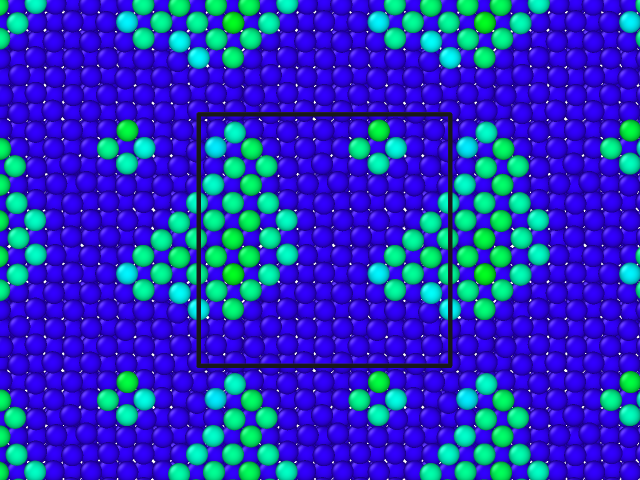}}
\subfigure[ t=4 ms (-2.6 eV)]{
\includegraphics[scale=0.118]{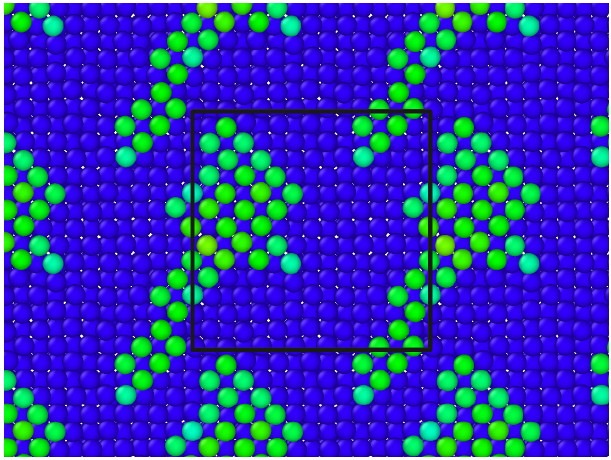}}
\subfigure[ t=9 ms (-3.7 eV)]{
\includegraphics[scale=0.118]{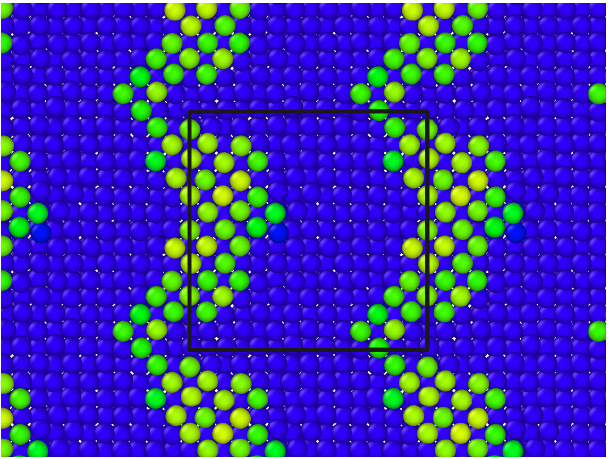}}
\subfigure[ t=15 ms(-5.0 eV)]{
\includegraphics[scale=0.118]{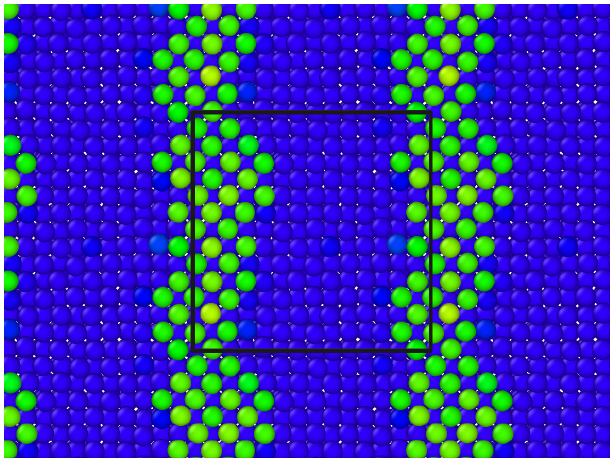}}
\caption{\label{fig:islandripening} Room-temperature adatom ripening on Al (001) surface as a function of time. Each structure was quenched to find its corresponding local minima, and the energies after quenching (relative to (a)) are provided. (a) shows the starting geometry. After 1$\mu$s, we have two disconnected clusters (b-c). Corresponding brute-force MD runs were found trapped in similar configuration in the fraction of microsecond time they could achieve. At around 4ms these two clusters join (d). At 9ms (e) there is further joining and we have effectively one long chain of adatoms. At 15ms (f) the chain has coarsened into one entity across simulation cells. Color scheme same as in Fig. 3. }
\end{figure}

Fig. \ref{fig:mechanisms} and the movies in SM illustrate the most common mechanisms seen through SISYPHUS. We have the single adatom hop (a-b), the concerted two adatom hop (c-d), and the concerted event involving an adatom and a substrate atom as the latter moves to the adatom layer (e-f). Occasionally we see more complicated mechanisms like the 3-atom hop (g-h), and events with creation of a vacancy in the top substrate layer (see SM). The first three mechanisms are the most common and are in fact the lowest energy transitions found using the dimer method for saddle point search. In Fig. \ref{fig:islandripening} and movie in SM we show the typical evolution of the island ripening at room temperature over several milliseconds of physical time (exact time can vary from run to run well but still within an order of magnitude). Shown alongwith are corresponding energies obtained by quenching each structure to its local minima, illustrating further the effectiveness of the algorithm in escaping and exploring various local energy minima. The overall time can be compared with the analogous on-the-fly KMC work\cite{ontheflykmc} for same system and interatomic potential\cite{voterchen} where 20 adatoms (half as many as current work) formed one compact cluster around 1ms.
As a verification that our proposed BDF is effective to decomposing phase space into disconnected wells, we show, in a movie accompanying the SM, a superposition of snapshots of the system during a MC$a$ run, illustrating that the system does not jump from one well to another within one MC$a$ run (since it rejects all moves with $\chi\geq\chi_{cut}$). 

In conclusion, we have shown SISYPHUS to be an extremely parallelizable and robust set of algorithms that help achieve fraction of second timescale for thousands of atoms.  The method works well irrespective of system size, and can be applied in the general setting of any collective variable. We also introduced a new CV appropriate for solid state systems especially for transitions involving collective motions of several atoms.

We would like to thank Dr. Arthur Voter and Prof. Graeme Henkelman for helpful discussions and stimulating the trajectory of this research, Dr. Christopher Weinberger for the HTST calculation for Ta vacancy diffusion, and Prof. Julia Greer for hospitality during PT's visit to Caltech. This research was supported by the US National Science Foundation through XSEDE computational resources provided by NCSA under grant DMR050013N and DMR120056, and NSF Condensed Matter and Materials Theory program DMR0907669, and the Office of Naval Research through grant N00014-11-1-0886.


\begin{thebibliography}{30}
\expandafter\ifx\csname natexlab\endcsname\relax\def\natexlab#1{#1}\fi
\expandafter\ifx\csname bibnamefont\endcsname\relax
  \def\bibnamefont#1{#1}\fi
\expandafter\ifx\csname bibfnamefont\endcsname\relax
  \def\bibfnamefont#1{#1}\fi
\expandafter\ifx\csname citenamefont\endcsname\relax
  \def\citenamefont#1{#1}\fi
\expandafter\ifx\csname url\endcsname\relax
  \def\url#1{\texttt{#1}}\fi
\expandafter\ifx\csname urlprefix\endcsname\relax\def\urlprefix{URL }\fi
\providecommand{\bibinfo}[2]{#2}
\providecommand{\eprint}[2][]{\url{#2}}

\bibitem[{\citenamefont{Laio and Parrinello}(2002)}]{metadynamics}
\bibinfo{author}{\bibfnamefont{A.}~\bibnamefont{Laio}} \bibnamefont{and}
  \bibinfo{author}{\bibfnamefont{M.}~\bibnamefont{Parrinello}},
  \bibinfo{journal}{Proceedings of the National Academy of Sciences}
  \textbf{\bibinfo{volume}{99}}, \bibinfo{pages}{12562} (\bibinfo{year}{2002}).

\bibitem[{\citenamefont{Voter}(1997)}]{hyperdynamics}
\bibinfo{author}{\bibfnamefont{A.~F.} \bibnamefont{Voter}},
  \bibinfo{journal}{Phys. Rev. Lett.} \textbf{\bibinfo{volume}{78}},
  \bibinfo{pages}{3908} (\bibinfo{year}{1997}).

\bibitem[{\citenamefont{Tiwary and van~de Walle}(2011)}]{mcmd}
\bibinfo{author}{\bibfnamefont{P.}~\bibnamefont{Tiwary}} \bibnamefont{and}
  \bibinfo{author}{\bibfnamefont{A.}~\bibnamefont{van~de Walle}},
  \bibinfo{journal}{Phys. Rev. B} \textbf{\bibinfo{volume}{84}},
  \bibinfo{pages}{100301} (\bibinfo{year}{2011}).

\bibitem[{\citenamefont{Fan et~al.}(2011)\citenamefont{Fan, Kushima, Yip, and
  Yildiz}}]{abc}
\bibinfo{author}{\bibfnamefont{Y.}~\bibnamefont{Fan}},
  \bibinfo{author}{\bibfnamefont{A.}~\bibnamefont{Kushima}},
  \bibinfo{author}{\bibfnamefont{S.}~\bibnamefont{Yip}}, \bibnamefont{and}
  \bibinfo{author}{\bibfnamefont{B.}~\bibnamefont{Yildiz}},
  \bibinfo{journal}{Phys. Rev. Lett.} \textbf{\bibinfo{volume}{106}},
  \bibinfo{pages}{125501} (\bibinfo{year}{2011}).

\bibitem[{\citenamefont{Henkelman and Jonsson}(2001)}]{ontheflykmc}
\bibinfo{author}{\bibfnamefont{G.}~\bibnamefont{Henkelman}} \bibnamefont{and}
  \bibinfo{author}{\bibfnamefont{H.}~\bibnamefont{Jonsson}},
  \bibinfo{journal}{The Journal of Chemical Physics}
  \textbf{\bibinfo{volume}{115}}, \bibinfo{pages}{9657} (\bibinfo{year}{2001}).

\bibitem[{\citenamefont{Barkema and Mousseau}(1996)}]{art}
\bibinfo{author}{\bibfnamefont{G.~T.} \bibnamefont{Barkema}} \bibnamefont{and}
  \bibinfo{author}{\bibfnamefont{N.}~\bibnamefont{Mousseau}},
  \bibinfo{journal}{Phys. Rev. Lett.} \textbf{\bibinfo{volume}{77}},
  \bibinfo{pages}{4358} (\bibinfo{year}{1996}).

\bibitem[{\citenamefont{Lindorff-Larsen
  et~al.}(2011)\citenamefont{Lindorff-Larsen, Piana, Dror, and Shaw}}]{deshaw}
\bibinfo{author}{\bibfnamefont{K.}~\bibnamefont{Lindorff-Larsen}},
  \bibinfo{author}{\bibfnamefont{S.}~\bibnamefont{Piana}},
  \bibinfo{author}{\bibfnamefont{R.~O.} \bibnamefont{Dror}}, \bibnamefont{and}
  \bibinfo{author}{\bibfnamefont{D.~E.} \bibnamefont{Shaw}},
  \bibinfo{journal}{Science} \textbf{\bibinfo{volume}{334}},
  \bibinfo{pages}{517} (\bibinfo{year}{2011}).

\bibitem[{\citenamefont{Lu et~al.}(2010)\citenamefont{Lu, Makarov, and
  Henkelman}}]{kappa}
\bibinfo{author}{\bibfnamefont{C.-Y.} \bibnamefont{Lu}},
  \bibinfo{author}{\bibfnamefont{D.~E.} \bibnamefont{Makarov}},
  \bibnamefont{and}
  \bibinfo{author}{\bibfnamefont{G.}~\bibnamefont{Henkelman}},
  \bibinfo{journal}{The Journal of Chemical Physics}
  \textbf{\bibinfo{volume}{133}}, \bibinfo{eid}{201101}
  (pages~\bibinfo{numpages}{4}) (\bibinfo{year}{2010}).

\bibitem[{\citenamefont{Bai et~al.}(2010)\citenamefont{Bai, Voter, Hoagland,
  Nastasi, and Uberuaga}}]{blas_science}
\bibinfo{author}{\bibfnamefont{X.-M.} \bibnamefont{Bai}},
  \bibinfo{author}{\bibfnamefont{A.~F.} \bibnamefont{Voter}},
  \bibinfo{author}{\bibfnamefont{R.~G.} \bibnamefont{Hoagland}},
  \bibinfo{author}{\bibfnamefont{M.}~\bibnamefont{Nastasi}}, \bibnamefont{and}
  \bibinfo{author}{\bibfnamefont{B.~P.} \bibnamefont{Uberuaga}},
  \bibinfo{journal}{Science} \textbf{\bibinfo{volume}{327}},
  \bibinfo{pages}{1631} (\bibinfo{year}{2010}).

\bibitem[{\citenamefont{Voter et~al.}(2002)\citenamefont{Voter, Montalenti, and
  Germann}}]{voter_review}
\bibinfo{author}{\bibfnamefont{A.}~\bibnamefont{Voter}},
  \bibinfo{author}{\bibfnamefont{F.}~\bibnamefont{Montalenti}},
  \bibnamefont{and} \bibinfo{author}{\bibfnamefont{T.}~\bibnamefont{Germann}},
  \bibinfo{journal}{Annual Review of Materials Research}
  \textbf{\bibinfo{volume}{32}}, \bibinfo{pages}{321} (\bibinfo{year}{2002}).

\bibitem[{\citenamefont{Laio and Gervasio}(2008)}]{laio_review}
\bibinfo{author}{\bibfnamefont{A.}~\bibnamefont{Laio}} \bibnamefont{and}
  \bibinfo{author}{\bibfnamefont{F.~L.} \bibnamefont{Gervasio}},
  \bibinfo{journal}{Reports on Progress in Physics}
  \textbf{\bibinfo{volume}{71}}, \bibinfo{pages}{126601}
  (\bibinfo{year}{2008}).

\bibitem[{\citenamefont{Perez et~al.}(2009)\citenamefont{Perez, Uberuaga, Shim,
  Amar, and Voter}}]{perez_review}
\bibinfo{author}{\bibfnamefont{D.}~\bibnamefont{Perez}},
  \bibinfo{author}{\bibfnamefont{B.}~\bibnamefont{Uberuaga}},
  \bibinfo{author}{\bibfnamefont{Y.}~\bibnamefont{Shim}},
  \bibinfo{author}{\bibfnamefont{J.}~\bibnamefont{Amar}}, \bibnamefont{and}
  \bibinfo{author}{\bibfnamefont{A.}~\bibnamefont{Voter}},
  \bibinfo{journal}{Annual Reports in Computational Chemistry}
  \textbf{\bibinfo{volume}{5}}, \bibinfo{pages}{79} (\bibinfo{year}{2009}).

\bibitem[{\citenamefont{S{\o}rensen and Voter}(2000)}]{tad}
\bibinfo{author}{\bibfnamefont{M.}~\bibnamefont{S{\o}rensen}} \bibnamefont{and}
  \bibinfo{author}{\bibfnamefont{A.}~\bibnamefont{Voter}},
  \bibinfo{journal}{The Journal of Chemical Physics}
  \textbf{\bibinfo{volume}{112}}, \bibinfo{pages}{9599} (\bibinfo{year}{2000}).

\bibitem[{\citenamefont{Jarzynski}(2006)}]{jarzynski_rare}
\bibinfo{author}{\bibfnamefont{C.}~\bibnamefont{Jarzynski}},
  \bibinfo{journal}{Physical Review E} \textbf{\bibinfo{volume}{73}},
  \bibinfo{pages}{046105} (\bibinfo{year}{2006}).

\bibitem[{\citenamefont{Bussi et~al.}(2006)\citenamefont{Bussi, Laio, and
  Parrinello}}]{metadynamics_energy}
\bibinfo{author}{\bibfnamefont{G.}~\bibnamefont{Bussi}},
  \bibinfo{author}{\bibfnamefont{A.}~\bibnamefont{Laio}}, \bibnamefont{and}
  \bibinfo{author}{\bibfnamefont{M.}~\bibnamefont{Parrinello}},
  \bibinfo{journal}{Phys. Rev. Lett.} \textbf{\bibinfo{volume}{96}},
  \bibinfo{pages}{090601} (\bibinfo{year}{2006}).

\bibitem[{\citenamefont{Tribello et~al.}(2010)\citenamefont{Tribello, Ceriotti,
  and Parrinello}}]{sketchmap1}
\bibinfo{author}{\bibfnamefont{G.~A.} \bibnamefont{Tribello}},
  \bibinfo{author}{\bibfnamefont{M.}~\bibnamefont{Ceriotti}}, \bibnamefont{and}
  \bibinfo{author}{\bibfnamefont{M.}~\bibnamefont{Parrinello}},
  \bibinfo{journal}{Proceedings of the National Academy of Sciences}
  \textbf{\bibinfo{volume}{107}}, \bibinfo{pages}{17509}
  (\bibinfo{year}{2010}).

\bibitem[{\citenamefont{Tribello et~al.}(2012)\citenamefont{Tribello, Ceriotti,
  and Parrinello}}]{sketchmap2}
\bibinfo{author}{\bibfnamefont{G.~A.} \bibnamefont{Tribello}},
  \bibinfo{author}{\bibfnamefont{M.}~\bibnamefont{Ceriotti}}, \bibnamefont{and}
  \bibinfo{author}{\bibfnamefont{M.}~\bibnamefont{Parrinello}},
  \bibinfo{journal}{Proceedings of the National Academy of Sciences}
  \textbf{\bibinfo{volume}{109}}, \bibinfo{pages}{5196} (\bibinfo{year}{2012}).

\bibitem[{\citenamefont{Voter}(1998)}]{parrep}
\bibinfo{author}{\bibfnamefont{A.~F.}~\bibnamefont{Voter}},
  \bibinfo{journal}{Physical Review B} \textbf{\bibinfo{volume}{57}},
  \bibinfo{pages}{13985} (\bibinfo{year}{1998}).

\bibitem[{\citenamefont{Frenkel and Smit}(2002)}]{frenkelsmit}
\bibinfo{author}{\bibfnamefont{D.}~\bibnamefont{Frenkel}} \bibnamefont{and}
  \bibinfo{author}{\bibfnamefont{B.}~\bibnamefont{Smit}},
  \emph{\bibinfo{title}{{Understanding molecular simulation : from algorithms
  to applications}}} (\bibinfo{publisher}{Academic Press},
  \bibinfo{year}{2002}), \bibinfo{edition}{2nd} ed., ISBN
  \bibinfo{isbn}{0122673514}.

\bibitem[{\citenamefont{Jarzynski}(1997)}]{jarzynski}
\bibinfo{author}{\bibfnamefont{C.}~\bibnamefont{Jarzynski}},
  \bibinfo{journal}{Phys. Rev. Lett.} \textbf{\bibinfo{volume}{78}},
  \bibinfo{pages}{2690} (\bibinfo{year}{1997}).

\bibitem[{\citenamefont{Fichthorn et~al.}(2009)\citenamefont{Fichthorn, Miron,
  Wang, and Tiwary}}]{bondboost}
\bibinfo{author}{\bibfnamefont{K.~A.} \bibnamefont{Fichthorn}},
  \bibinfo{author}{\bibfnamefont{R.~A.} \bibnamefont{Miron}},
  \bibinfo{author}{\bibfnamefont{Y.}~\bibnamefont{Wang}}, \bibnamefont{and}
  \bibinfo{author}{\bibfnamefont{Y.}~\bibnamefont{Tiwary}},
  \bibinfo{journal}{Journal of Physics: Condensed Matter}
  \textbf{\bibinfo{volume}{21}}, \bibinfo{pages}{084212}
  (\bibinfo{year}{2009}).

\bibitem[{\citenamefont{Falk and Langer}(1998)}]{stz}
\bibinfo{author}{\bibfnamefont{M.~L.} \bibnamefont{Falk}} \bibnamefont{and}
  \bibinfo{author}{\bibfnamefont{J.~S.} \bibnamefont{Langer}},
  \bibinfo{journal}{Phys. Rev. E} \textbf{\bibinfo{volume}{57}},
  \bibinfo{pages}{7192} (\bibinfo{year}{1998}).

\bibitem[{\citenamefont{Mrowec and Marcinkiewicz}(1980)}]{defectsmaterials}
\bibinfo{author}{\bibfnamefont{S.}~\bibnamefont{Mrowec}} \bibnamefont{and}
  \bibinfo{author}{\bibfnamefont{S.}~\bibnamefont{Marcinkiewicz}},
  \emph{\bibinfo{title}{Defects and diffusion in solids: an introduction}}
  (\bibinfo{publisher}{Elsevier}, \bibinfo{year}{1980}).

\bibitem[{\citenamefont{Watson and Baxter}(2007)}]{epsl}
\bibinfo{author}{\bibfnamefont{E.~B.} \bibnamefont{Watson}} \bibnamefont{and}
  \bibinfo{author}{\bibfnamefont{E.~F.} \bibnamefont{Baxter}},
  \bibinfo{journal}{Earth and Planetary Science Letters}
  \textbf{\bibinfo{volume}{253}}, \bibinfo{pages}{307} (\bibinfo{year}{2007}).

\bibitem[{\citenamefont{Mendelev and Mishin}(2009)}]{mendelev}
\bibinfo{author}{\bibfnamefont{M.~I.} \bibnamefont{Mendelev}} \bibnamefont{and}
  \bibinfo{author}{\bibfnamefont{Y.}~\bibnamefont{Mishin}},
  \bibinfo{journal}{Phys. Rev. B} \textbf{\bibinfo{volume}{80}},
  \bibinfo{pages}{144111} (\bibinfo{year}{2009}).

\bibitem[{\citenamefont{Ackland and Thetford}(1987)}]{ackland}
\bibinfo{author}{\bibfnamefont{G.}~\bibnamefont{Ackland}} \bibnamefont{and}
  \bibinfo{author}{\bibfnamefont{R.}~\bibnamefont{Thetford}},
  \bibinfo{journal}{Philosophical Magazine A} \textbf{\bibinfo{volume}{56}},
  \bibinfo{pages}{15} (\bibinfo{year}{1987}).

\bibitem[{\citenamefont{Zhang and Lagally}(1997)}]{thinfilm_science}
\bibinfo{author}{\bibfnamefont{Z.}~\bibnamefont{Zhang}} \bibnamefont{and}
  \bibinfo{author}{\bibfnamefont{M.~G.} \bibnamefont{Lagally}},
  \bibinfo{journal}{Science} \textbf{\bibinfo{volume}{276}},
  \bibinfo{pages}{377} (\bibinfo{year}{1997}).

\bibitem[{\citenamefont{Becker et~al.}(2009)\citenamefont{Becker, Mignogna, and
  Fichthorn}}]{fichthorn_prl}
\bibinfo{author}{\bibfnamefont{K.~E.} \bibnamefont{Becker}},
  \bibinfo{author}{\bibfnamefont{M.~H.} \bibnamefont{Mignogna}},
  \bibnamefont{and} \bibinfo{author}{\bibfnamefont{K.~A.}
  \bibnamefont{Fichthorn}}, \bibinfo{journal}{Phys. Rev. Lett.}
  \textbf{\bibinfo{volume}{102}}, \bibinfo{pages}{046101}
  (\bibinfo{year}{2009}).

\bibitem[{\citenamefont{Henkelman and Jonsson}(1999)}]{dimer}
\bibinfo{author}{\bibfnamefont{G.}~\bibnamefont{Henkelman}} \bibnamefont{and}
  \bibinfo{author}{\bibfnamefont{H.}~\bibnamefont{Jonsson}},
  \bibinfo{journal}{The Journal of Chemical Physics}
  \textbf{\bibinfo{volume}{111}}, \bibinfo{pages}{7010} (\bibinfo{year}{1999}).

\bibitem[{\citenamefont{Voter and Chen}(1987)}]{voterchen}
\bibinfo{author}{\bibfnamefont{A.}~\bibnamefont{Voter}} \bibnamefont{and}
  \bibinfo{author}{\bibfnamefont{S.}~\bibnamefont{Chen}}, in
  \emph{\bibinfo{booktitle}{Mater. Res. Soc. Proc}} (\bibinfo{year}{1987}),
  vol.~\bibinfo{volume}{82}, pp. \bibinfo{pages}{175--180}.

\end{thebibliography}
\end{document}